\documentclass[12pt]{article}
\usepackage[dvips]{graphics,color}
\usepackage{amssymb}
\usepackage{amsmath}

\def \IC{\mathbb{C}}
\def \IZ{\mathbb{Z}}
\def \IP{\mathbb{P}}

\def \tr{\mbox{Tr}}

\def\ba{\begin{eqnarray}}
\def\ea{\end{eqnarray}}
\def\W{{\cal W}}
\def\O{{\cal O}}
\def\sf{s_{\tableau{1}}}

\def \ao{\alpha_1}
\def \at{\alpha_2}

\def \bo{\beta_1}

\def \att{\alpha_2^t}

\def \bot{\beta_1^t}

\def \Qn{\tilde Q}
\def \N{{\cal N}}
\def \nn{\nonumber}
\def \K{{\cal K}}

\def \[{[ \hspace{-0.1cm} \{}
\def \]{\} \hspace{-0.1cm} ]}

\newdimen\tableauside\tableauside=1.0ex
\newdimen\tableaurule\tableaurule=0.4pt
\newdimen\tableaustep
\def\phantomhrule#1{\hbox{\vbox to0pt{\hrule height\tableaurule width#1\vss}}}
\def\phantomvrule#1{\vbox{\hbox to0pt{\vrule width\tableaurule height#1\hss}}}
\def\sqr{\vbox{%
  \phantomhrule\tableaustep
  \hbox{\phantomvrule\tableaustep\kern\tableaustep\phantomvrule\tableaustep}%
  \hbox{\vbox{\phantomhrule\tableauside}\kern-\tableaurule}}}
\def\squares#1{\hbox{\count0=#1\noindent\loop\sqr
  \advance\count0 by-1 \ifnum\count0>0\repeat}}
\def\tableau#1{\vcenter{\offinterlineskip
  \tableaustep=\tableauside\advance\tableaustep by-\tableaurule
  \kern\normallineskip\hbox
    {\kern\normallineskip\vbox
      {\gettableau#1 0 }%
     \kern\normallineskip\kern\tableaurule}%
  \kern\normallineskip\kern\tableaurule}}
\def\gettableau#1 {\ifnum#1=0\let\next=\null\else
  \squares{#1}\let\next=\gettableau\fi\next}

\begin{document}
\noindent
\begin{titlepage}

\begin{center}
\today \hfill SMS-0402\\\hfill SLAC-PUB-10804 \\
\hfill SU-ITP 4/39 \\  \hfill hep-th/0410174\\ \vskip 1cm {\large {\bf
The Vertex on a Strip}} \vskip 2cm {Amer Iqbal$^{\spadesuit}$,
Amir-Kian Kashani-Poor$^{\clubsuit}$}\\ \vskip 0.5cm

{$^{\spadesuit}$School of Mathematical Sciences\\
GC University,\\
Lahore, 54600, Pakistan.\\} \vskip 0.5cm

{$^{\clubsuit}$ Department of Physics and SLAC\\
Stanford University,\\
Stanford, CA 94305/94309, U.S.A.\\}\vskip 0.5cm

{$^{\spadesuit}$ Department of Mathematics\\
University of Washington,\\
Seattle, WA, 98195, U.S.A.\\}

\end{center}

\begin{abstract}
We demonstrate that for a broad class of local Calabi-Yau geometries built around a string of $\IP^1$'s -- those whose toric diagrams are given by triangulations of a strip -- we can derive simple rules, based on the topological vertex, for obtaining expressions for the topological string partition function in which the sums over Young tableaux have been performed. By allowing non-trivial tableaux on the external legs of the corresponding web diagrams, these strips can be used as building blocks for more general geometries. As applications of our result, we study the behavior of topological string amplitudes under flops, as well as check Nekrasov's conjecture in its most general form.
\end{abstract}
\end{titlepage}
\newpage

\section{Introduction}
In the last few years, dramatic progress has been made in techniques for calculating the partition function of the topological string on toric (hence non-compact) Calabi-Yau manifolds \cite{Diaconescu:2002sf,Diaconescu:2002qf,Aganagic:2002qg}. The culmination of this effort has been the formulation of the topological vertex \cite{Aganagic:2003db} (see \cite{Li} for a recent mathematical treatment). With it, a set of diagrammatic rules can be formulated which allow an expression for the topological string partition function to be read off from the web diagram of the toric manifold. While the expressions obtained such are algorithmically complete, they contain unwieldy sums over Young tableaux, one sum for each internal line of the web diagram. Starting with \cite{Iqbal:2003ix}, methods were developed to perform a portion of these sums \cite{Iqbal:2003zz,Eguchi:2003sj,Zhou:2003zp,Eguchi:2003it}. In this note, we show how to perform all sums which arise in an arbitrary smooth triangulation of a strip toric diagram, such as
\begin{figure}[h]
\begin{center}\leavevmode\includegraphics{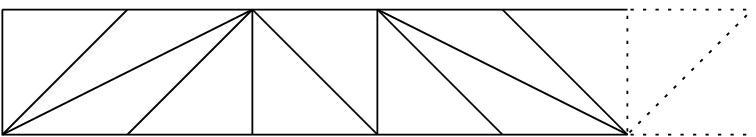}\end{center}
\end{figure}

\vspace{-1cm}
with arbitrary representations on all external legs but the first and last.

The ultimate goal of this program is to provide a technique for efficiently extracting the Gopakumar-Vafa invariants from the expressions the topological vertex yields for the topological string partition function. We will outline the obstacles to this goal using the methods of this paper as we proceed. 

As other applications, we offer an analysis of the behavior of the topological amplitude under flops of the target manifold. We demonstrate that the Gopakumar-Vafa invariants for all toric geometries decomposable into strips are invariant under flops. We also show that our results provide the framework to check Nekrasov's results \cite{Nekrasov:2003af} in the most general case of product $U(N)$ gauge groups with any number of allowed hypermultiplets.

The organization of this paper is as follows. In section 2, we elucidate the geometries we are considering and present and interpret the rules for obtaining the topological string partition function on them. In section  3, we derive these results. We include a brief review of the topological vertex at the beginning of this section, and end it with a comparison to the natural 4-vertex obtained from Chern-Simons theory. We discuss the behavior of Gopakumar-Vafa invariants under flops on geometries decomposable into strips in section 4.1. Section 4.2 provides the basic building blocks to study Nekrasov's conjecture. We end with conclusions. An appendix gives a brief introduction to Schur functions, and collects the identities for Schur functions used throughout the paper.

\section{The results}
\subsection{Geometry of the strip}
Recall that a simple way of visualizing the geometries given by toric diagrams is to think of them as $T^n$ fibrations over $n$ dimensional base manifolds with corners (see eg. section 4.1 of \cite{fulton93}). Locally, one can introduce complex coordinates on the toric manifold. The base manifold is then locally given by the absolute value of these coordinates, the $T^n$ by the phases. The boundary of the base is where some of these coordinates vanish, entailing a degeneration of the corresponding number of fiber directions.

\begin{figure}[h]
\begin{center}\leavevmode\includegraphics{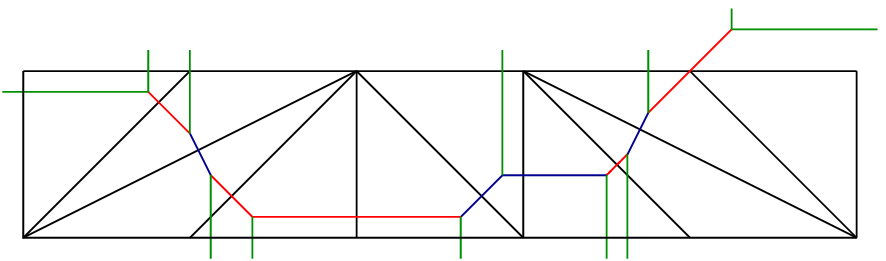}\end{center}
\caption{\small Relation between toric and web diagram. \label{torictoweb}}
\end{figure}

In 3 complex dimensions, the 3 real dimensional base has a 2 dimensional boundary with edges and corner. Web diagrams, easily obtained from toric diagrams as sketched in figure \ref{torictoweb}, represent the projection of the edges and corners of the base on to the plane. There is a full $T^3$ fibered over each point above the plane, corresponding to the interior of the base manifold. On a generic point on the plane, representing a generic point on the boundary, one cycle of the fiber degenerates, two degenerate on the lines of the web diagram, which correspond to edges of the base, and the entire fiber degenerates at the vertices of the diagram, the corners of the base.

Returning to figure \ref{torictoweb}, we now see the string of $\IP^1$'s (in red and blue) emerging by following the $S^1$ fibration along the internal line running through the web diagram. It is capped off to $\IP^1$'s by the $S^1$'s degenerating at each vertex. The two non-compact directions of the geometry locally correspond to the sum over two line bundles over each $\IP^1$. The two local geometries that arise on the strip are $(\O(-2) \oplus \O) \rightarrow \IP^1$ (in red) and $(\O(-1) \oplus \O(-1))\rightarrow \IP^1$ (in blue). We refer to the respective $\IP^1$'s as $(-2,0)$ and $(-1,-1)$ curves in the following.

\subsection{Rules on the strip}
Each vertex has one non-trivial Young tableau associated to it, the two outer vertices in addition have one leg carrying the trivial tableau. All other indices of the vertices are summed over. We label the non-trivial tableaux by $\beta_i$, with $i$ indexing the vertex. The internal lines carry a factor $Q_i = e^{-t_i}$, where $t_i$ is the K\"ahler parameter of the curve the internal line represents.

\begin{itemize}
\item{
Each vertex contributes a factor of $\W_{\beta_i}=s_{\beta_i}(q^\rho)$ (the notation for the argument of the Schur function is explained in the next section).}
\item{Each pair of vertices (not just adjacent ones) contributes a factor to the amplitude, which is a pairing of the non-trivial tableaux carried by the pair. The interpretation of this observation is that branes wrapping the curves consisting of touching $\IP^1$'s in the web diagram contribute to the Gopakumar-Vafa index just as those wrapping the individual $\IP^1$'s.}
\item{While the pairing itself is symmetric, for the purpose of book keeping, we will choose one of the two natural orderings of the vertices along the string of $\IP^1$'s. We will speak of the first or second slot of the pairing with reference to this ordering. To determine the pairing factor, note that two types of curves occur on the strip: $(-2,0)$ curves and $(-1,-1)$ curves. Up to $SL(2,\IZ)$ transformations, these are represented by the toric/web diagrams depicted in figure 
\ref{twogoodlinks}.
\begin{figure}[h]
\begin{center}
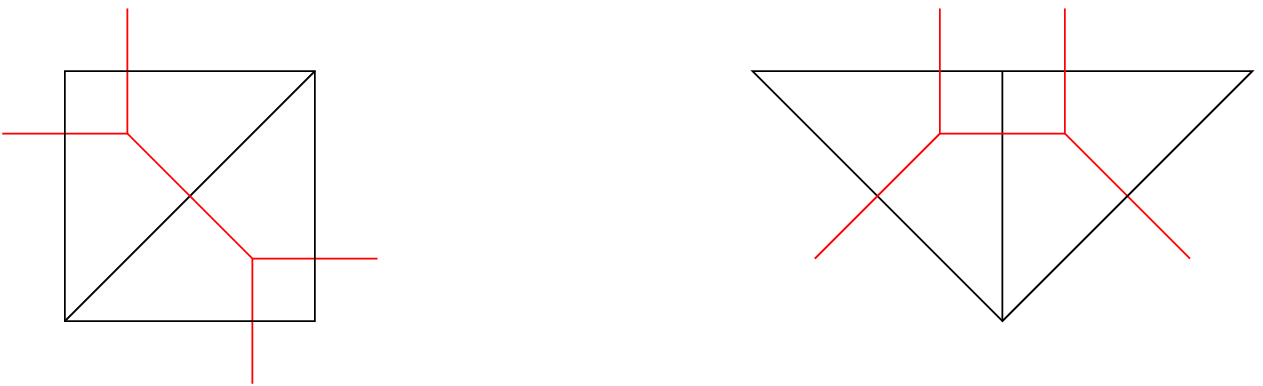
\end{center}
\caption{\small The two building blocks of a webdiagram on the strip. \label{twogoodlinks}}
\end{figure}
The contribution of the pairing to the amplitude depends essentially on whether an even or odd number of $(-1,-1)$ curves lie between the two vertices. The geometric interpretation of this fact is that the curves consisting of touching $\IP^1$'s in the web diagram have normal bundle $\O(-2) \oplus \O$ or $\O(-1) \oplus \O(-1)$, depending on whether the string of $\IP^1$'s contains an even or odd number of smooth $(-1,-1)$ curves. This is suggested both by the toric diagram and by the expressions for the two pairings, as we will see next.}
\item{We denote the two types of pairings between vertices carrying the Young tableaux $\alpha$ and $\beta$ as $\{ \alpha \beta\}$ and $[\alpha \beta ]=\{\alpha \beta\}^{-1}$ (we use the notation $\[ \alpha \beta \]$ when we make statements valid for both types of pairing).
The pairing $\{\alpha \beta\}$ is given by the expression
\ba
\{\alpha \beta\}_Q &=&  \prod_k  (1- Q q^k)^{C_k(\alpha, \beta)} \exp \left[ \sum_{n=1}^\infty \frac{Q^n}{n(2 \sin( \frac{n g_s}{2}))^2} \right] \,.
\ea
The product over $k$ is over a finite range of integers (possibly negative), $C_k(\alpha,\beta)$ are numbers which depend on the two Young tableaux that are being paired, given by
\ba 
\sum_k C_k(\alpha, \beta)q^k &=& \frac{q}{(q-1)^2} \left( 1 + (q-1)^2 \sum_{i=1}^{d_\alpha} q^{-i} \sum_{j=0}^{\alpha_i-1} q^j \right) \left( 1 + (q-1)^2 \sum_{i=1}^{d_\beta} q^{-i} \sum_{j=0}^{\beta_i-1} q^j \right) \nonumber \\
& & - \frac{q}{(1-q)^2}  \,.
\ea
The factor $Q$ is the product of all $Q_i$ labeling the internal lines connecting the two vertices. Note that, as advertised above, if we take the two Young tableaux to be trivial, the contribution from the pairing $\{\cdot \cdot\}$ is exactly that of a $(-1,-1)$ curve, and likewise, the contribution of $[\cdot \cdot]=\{\cdot \cdot\}^{-1}$ is that of a $(-2,0)$ curve.}
\item{To keep track of the contribution from two paired vertices, we can divide the vertices into two relative types, $A$ and $B$, such that the type of a vertex depends on that of the preceding vertex: two vertices connected by a $(-2,0)$ curve are of same type, two connected by a $(-1,-1)$ curve of opposite type. If the vertices $i$ and $j$ are of same type (i.e. have an even number of $(-1,-1)$ curves between them), the pairing factor is $[\beta_i^{\cdot} \beta_j^{\cdot}]$, else $\{\beta_i^{\cdot} \beta_j^{\cdot} \} = [\beta_i^{\cdot} \beta_j^{\cdot}]^{-1} $.}
\item{The upper case dot indicates that either $\beta$ or $\beta^t$ is the correct entry. Either all pairings involving $\beta_i$ are of the form $\[ \beta_i \cdot \]$ and $\[ \cdot \beta_i^t \]$, or they are of the form $\[\beta_i^t \cdot\]$ and $\[\cdot \beta_i\]$. To determine which of the two options apply to $\beta_i$ for each $i$, we anchor the relative types $A$ and $B$ as follows: we will take the first vertex of the string of $\IP^1$'s to be of type $A$ if, labeling the legs in clockwise order, it is given by $C_{\alpha_1 \bullet \beta_1}$. Otherwise, it must be given by $C_{\bullet \alpha_1 \beta_1}$, and we will classify it as type $B$. With this convention, 
\ba
\mbox{$i$-th vertex of type $A$} &\leftrightarrow& \[\beta_i \cdot\] \,\mbox{and}\, \[\cdot \beta_i^t\] \,, \nonumber \\
\mbox{$i$-th vertex of type $B$} &\leftrightarrow& \[\beta_i^t \cdot \] \,\mbox{and}\, \[\cdot \beta_i\] \,. \nonumber
\ea}
\end{itemize}

\begin{figure}
\begin{center}
\resizebox{6cm}{!}{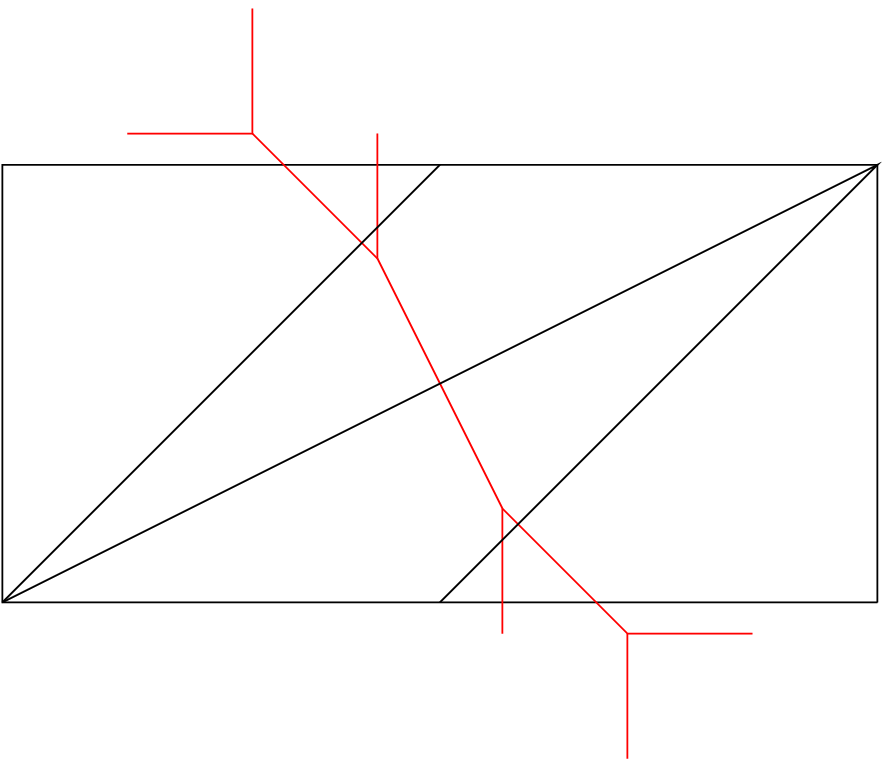}
\end{center}
\caption{\small A possible triangulation of a strip of length two. \label{example}}
\end{figure}

As an example, consider the diagram in figure \ref{example}. 
Starting from the left, the curves are of type $(-2,0)$, $(-1,-1)$, $(-2,0)$. The first vertex is $C_{\alpha_1 \bullet \beta_1}$, hence of type $A$. This determines the sequence of vertices to be $(A,A,B,B)$. By the rules above, we now obtain the following expression for the amplitude,
\ba
s_{\beta_1} s_{\beta_2} s_{\beta_3} s_{\beta_4} [\beta_1 \beta_2^t]_{Q_1} \{\beta_1 \beta_3 \}_{Q_1 Q_2} \{\beta_1 \beta_4 \}_{Q_1 Q_2 Q_3} \{\beta_2 \beta_3\}_{Q_2} \{\beta_2 \beta_4\}_{Q_2 Q_3} [\beta_3^t \beta_4]_{Q_3} =   \\
s_{\beta_1} s_{\beta_2} s_{\beta_3} s_{\beta_4} \frac{\{\beta_1 \beta_3 \}_{Q_1 Q_2} \{\beta_1 \beta_4 \}_{Q_1 Q_2 Q_3} \{\beta_2 \beta_3\}_{Q_2} \{\beta_2 \beta_4\}_{Q_2 Q_3}}{\{\beta_1 \beta_2^t\}_{Q_1} \{\beta_3^t \beta_4\}_{Q_3}} \;, \nn
\ea
where we have omitted the arguments $q^\rho$ of the Schur functions.

\section{Derivation}
\subsection{Review of the vertex}
Locally, any complex manifold is isomorphic to $\IC^n$. The topology and complex structure of the manifold are obtained by specifying how  these $\IC^n$-patches are to be glued together. The insight underlying the topological vertex \cite{Aganagic:2003db} is that the topological string partition function on a toric CY can also be pieced together patchwise. The patching conditions are implemented by placing non-compact Lagrangian D-branes along the three legs of the web diagram of $\IC^3$, intersecting the curves extending along these legs (recall that the legs indicate where 2 of the 3 cycles of the $T^3$ fibration have degenerated) in $S^1$'s. The topological string on each such patch counts the holomorphic curves ending on the branes, weighted by the appropriate Wilson lines from the boundaries of the worldsheet,
\ba \label{zpatch}
Z_{patch} &=& \sum_{\vec{k}^{(1)},\vec{k}^{(2)},\vec{k}^{(3)}} C_{\vec{k}^{(1)}\vec{k}^{(2)}\vec{k}^{(3)}} \times \mbox{Wilson loop factors}  \,.
\ea
The  vectors $\vec{k}^{(i)}$ encode that $k_j^{(i)}$ holes of winding number $j$ are ending on the $i$-th brane. The Wilson loop factors are given by
\ba
\mbox{Wilson loop factors} &=& \prod_{i=1}^{3} \frac{1}{\prod_j k_j^{(i)}!j^{k_j^{(i)}}} \prod_{j=1}^\infty (\tr V^j)^{k_j^{(i)}} \,.
\ea
The trace here is taken in the fundamental representation.
The definition of the vertex we will use in the following arises when rewriting (\ref{zpatch}) in the representation basis \cite{Aganagic:2003db},
\ba
\sum_{\alpha_1, \alpha_2, \alpha_3} C_{\alpha_1 \alpha_2 \alpha_3} \prod_{i=1}^{3} \tr_{\alpha_i} V_i &=& \sum_{\vec{k}^{(1)},\vec{k}^{(2)},\vec{k}^{(3)}} C_{\vec{k}^{(1)}\vec{k}^{(2)}\vec{k}^{(3)}} \prod_{i=1}^{3} \frac{1}{\prod_j k_j^{(i)}!j^{k_j^{(i)}}} \prod_{j=1}^\infty (\tr V^j)^{k_j^{(i)}} \,, \label{repbasis}
\ea
where the $\alpha_i$ now denote Young tableaux. This equation is to be understood in the limit when the number of D-branes on each leg is taken to infinity, such that the sum extends over Young tableaux with an arbitrary number of rows. An application of the Frobenius formula lets us solve (\ref{repbasis}) for $C_{\alpha_1 \alpha_2 \alpha_3}$.

The non-compactness of the Lagrangian D-branes gives rise to an integer ambiguity \cite{Aganagic:2001nx,Katz:2001vm}, which necessitates specifying one integer per leg to full determine the vertex. \cite{Aganagic:2003db} refer to this choice as the framing of the vertex, since the integer ambiguity  maps to the framing ambiguity of Chern-Simons theory under geometric transitions. The vertex in canonical framing is given by \cite{Okounkov:2003sp}
\begin{eqnarray}
C_{\lambda \mu \nu} = q^{\frac{\kappa(\lambda)}{2}}s_{\nu}(q^\rho)\sum_\eta s_{\lambda^t/\eta}(q^{\nu+\rho})s_{\mu/\eta}(q^{\nu^t+\rho}) \,.
\end{eqnarray}
The notation is $s(q^{\nu+\rho}) = s(\{q^{\nu_i -i +\frac{1}{2}}\})$. The $s_{\mu / \eta}$ are skew Schur functions, defined by
\ba \label{skew}
s_{\mu / \eta} = \sum_\nu c^{\mu}_{\eta \nu} s_\nu  \,,
\ea
where the $c^{\mu}_{\eta \nu}$ are tensor product coefficients, and $\kappa(\lambda) = \sum \lambda_i(\lambda_i - 2i+1)$. In the following, we will use the abbreviated notation $s_{\mu/\eta}(q^{\nu+\rho}) = \frac{\mu}{\eta}(q^{\nu+\rho})=\frac{\mu}{\eta}(\nu)$ whenever convenient, and imply a sum over repeated tableaux.

A framing must be specified for each leg of the vertex. If we represent each leg by an integer vector $v$, we can encode the framing by an integer vector $f$ that satisfies $f \wedge v =1$. The notation $f \wedge v$ denotes the symplectic product $f_1 v_2 -f_2 v_1$. The condition $f \wedge v =1$ determines $f$ up to integer multiples of $v$. Having chosen a canonical framing, we can hence classify relative framing by an integer $n$. To this end, we label the legs of the vertex in counter-clockwise order by $v_1, v_2, v_3$, s.t. $v_{i} \wedge v_{i+1}=1$. The natural choice for a framing is then $(f_1,f_2,f_3)=(v_3,v_1,v_2)$. Given a framing $f_i=v_{i-1}-n v_{i}$, the integer $n$ (the framing relative to the fiducial choice) is determined via $n = f_i \wedge v_{i-1}$. Under shifts of framing, the vertex transforms as follows \cite{Aganagic:2003db},
\ba
C_{\alpha_1 \alpha_2 \alpha_3}^{f_1 - n_1 v_1,f_2 - n_2 v_2,f_3 - n_3 v_3} &=& (-1)^{\sum_i n_i |\alpha_i|} q^{\sum_i n_i \frac{\kappa_{\alpha_i}}{2}} C_{\alpha_1 \alpha_2 \alpha_3}^{f_1,f_2,f_3}  \,.
\ea
Gluing two vertices together along $v_1$ and $v_1^\prime$ requires the framings along this leg to be opposite.\footnote{Thanks to Andy Neitzke for a discussion on this point that lead to the correction of a sign error in a previous version of this paper.} If we are gluing along $v_1$, and have canonical framing $f_1=v_3$ along $v_1$ for the first vertex, then the second vertex must have non-canonical framing $-v_3= v_3^\prime - n v_1^\prime$. We thus obtain the gluing rule
\ba
\sum_{\alpha_1} C_{\alpha_2 \alpha_3 \alpha_1} e^{-|\alpha_1| t} (-1)^{ |\alpha_1|}  C_{\alpha_1^t \alpha_2^\prime \alpha_3^\prime}^{-f_1,f_2^\prime,f_3^\prime}&=&\sum_{\alpha_1} C_{\alpha_2 \alpha_3 \alpha_1} e^{-|\alpha_1| t} (-1)^{ |\alpha_1|}  C_{\alpha_1^t \alpha_2^\prime \alpha_3^\prime}^{-v_3,f_2^\prime,f_3^\prime} \\
&=&\sum_{\alpha_1} C_{\alpha_2 \alpha_3 \alpha_1} e^{-|\alpha_1| t} (-1)^{|\alpha_1|}  C_{\alpha_1^t \alpha_2^\prime \alpha_3^\prime}^{v_3^\prime-n v_1^\prime,f_2^\prime,f_3^\prime} \nn \\
&=&\sum_{\alpha_1} C_{\alpha_2 \alpha_3 \alpha_1} e^{-|\alpha_1| t} (-1)^{(n+1) |\alpha_1|} q^{-n \kappa_{\alpha_1}/2} C_{\alpha_1^t \alpha_2^\prime \alpha_3^\prime}^{f_1^\prime,f_2^\prime,f_3^\prime}  \,, \nn
\ea
with $n= v_3^\prime \wedge (v_3^\prime + v_3)  = v_3^\prime \wedge v_3$.

\subsection{Performing the sums} 
In performing the sums, we will make use of the following two identities for skew Schur polynomials \cite{Macdonald},
\ba
\sum_\alpha s_{\alpha/\eta_1}(x)s_{\alpha/\eta_2}(y)&=&  \prod_{i,j} (1-x_i y_j)^{-1} \sum_\kappa s_{\eta_2/\kappa}(x)s_{\eta_1/\kappa}(y) \,, \label{sumruleone}\\
\sum_\alpha s_{\alpha^t/\eta_1}(x)s_{\alpha/\eta_2}(y) &=&  \prod_{i,j} (1+x_i y_j) \sum_\kappa s_{\eta_2^t/\kappa^t}(x)s_{\eta_1^t/\kappa}(y) \,.\label{sumruletwo}
\ea 
In the abbreviated notation introduced above, these sum rules become
\ba
\frac{\alpha}{\eta_1}(x)\frac{\alpha}{\eta_2}(y) &=& \prod_{i,j} (1-x_i y_j)^{-1}  \frac{\eta_2}{\kappa}(x)\frac{\eta_1}{\kappa}(y) \,, \\
\frac{\alpha^t}{\eta_1}(x)\frac{\alpha}{\eta_2}(y) &=& \prod_{i,j} (1+x_i y_j)  \frac{\eta_2^t}{\kappa^t}(x)\frac{\eta_1^t}{\kappa}(y) \,.
\ea
In the following, it will be convenient to rewrite the infinite products as follows,
\ba
\prod (1-x_i y_j)^{-1}  &=& \prod \exp \left[- \log (1 - x_i y_j) \right] \\
&=& \exp \left[  \sum_n \frac{1}{n} \sum_i x_i^n \sum_j y_j^n \right] \nn \\
&=& \exp \left[  \sum_n \frac{1}{n} \sf(x^n) \sf(y^n)  \right] \,,\nn \\
\prod (1+x_i y_j) &=& \exp \left[ - \sum_n \frac{(-1)^n}{n} \sf(x^n) \sf(y^n)  \right] \,.
\ea

Aside from the skew Schur functions, the expression for the topological vertex contains two additional elements: powers of the exponential of the K\"ahler parameters, $(\pm Q)^{|\alpha|}$, and powers of the exponential of the string coupling, $q^{\kappa(\alpha)}$. With recourse to the homogeneity property of the skew Schur polynomials, the former is easy to deal with. The latter poses a greater difficulty and is at the root of our calculations being confined to the strip. We hope to return to this difficulty in forthcoming work. For the two types of pairing occurring on the strip, the dependence on this factor cancels, as we demonstrate next.

\begin{figure}[h]
\begin{center}
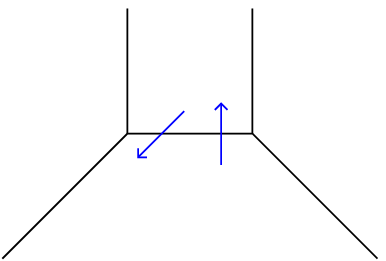
\end{center}
\caption{\small (-2,0) curve. \label{-2}}
\end{figure}

\paragraph{Type $[\beta_i \beta_j]$: (-2,0) curves} Consider figure \ref{-2}.
According to the rules reviewed above, the corresponding partition function is
\ba
\sum_{\alpha_2} C_{\alpha_2 \alpha_1 \beta_1} C_{\alpha_3 \alpha_2^t \beta_2}  (-1)^{|\alpha_2|}Q^{|\alpha_2|} (-1)^{n |\alpha_2|} q^{-\frac{n \kappa(\alpha_2)}{2}}  \,.
\ea
The $n$-dependence arises from the framing factors, where $n$ is given by 
\ba
n=v_{\beta_2} \wedge v_{\alpha_1} = (0,1)\wedge (-1,-1) = 1 \,.
\ea
Inserting the expression for the vertex, this yields
\ba \label{exprlink}
\sum_{\alpha_2} \bigg[ q^{\frac{\kappa(\alpha_2)}{2}}  \beta_1 \sum_{\eta_1} \frac{\alpha_2^t}{\eta_1}(\beta_1) \frac{\alpha_1}{\eta_1}(\beta_1^t) \bigg]  Q^{|\alpha_2|} q^{-\frac{\kappa(\alpha_2)}{2}} \\
 \bigg[q^{\frac{\kappa(\alpha_3)}{2}} \beta_2(q^\rho) \sum_{\eta_2} \frac{\alpha_3^t}{\eta_2}(\beta_2) \frac{\alpha_2^t}{\eta_2}(\beta_2^t) \bigg] \,. \nn
\ea
We see that the $q^{\kappa(\alpha_2)}$ dependence of the vertex cancels against the framing factor, so that we can apply (\ref{sumruleone}) to perform the sum over $\alpha_2$, once we deal with the $Q^{|\alpha_2|}$ dependence. As alluded to above, this does not pose any difficulty due to the homogeneity of the Schur functions, $s_{\lambda}(c\,x) = c^{|\lambda|} s_{\lambda}(x)$. Since the tensor product coefficients $c^{\mu}_{\eta \nu}$ vanish unless $|\mu|=|\eta|+|\nu|$, we easily deduce the homogeneity property of the skew Schur polynomials to be $s_{\mu/\eta}(c\,x) =\sum_{\nu} c^{\mu}_{\eta \nu} s_{\nu} (c\,x) = c^{|\mu|-|\eta|}s_{\mu/\eta}(x)$. We can now incorporate the K\"ahler parameters into our calculation. We obtain
\ba
\sum_{\alpha} \frac{\alpha}{\eta_1}(\beta_1) \frac{\alpha}{\eta_2}(\beta_2) Q^{|\alpha|} &=&  \sum_{\alpha} \frac{\alpha}{\eta_1}(q^{\rho+\beta_1}Q) \frac{\alpha}{\eta_2}(q^{\rho+\beta_2}) Q^{|\eta_1|} \\
&=& \exp\left[\sum_n \frac{1}{n} \sf\left((q^{\rho+\beta_1}Q)^n\right)\sf\left((q^{\rho+\beta_2})^n\right)\right] \sum_{\kappa} \frac{\eta_2}{\kappa} (q^{\rho+\beta_1}Q) \frac{\eta_1}{\kappa}(q^{\rho+\beta_2}) Q^{|\eta_1|}  \nonumber \\
&=&   \exp\left[\sum_n \frac{Q^n}{n} \sf\left((q^{\rho+\beta_1})^n\right)\sf\left((q^{\rho+\beta_2})^n\right)\right] \nn \\
& &\sum_{\kappa} \frac{\eta_2}{\kappa} (q^{\rho+\beta_1}) \frac{\eta_1}{\kappa}(q^{\rho+\beta_2}) Q^{|\eta_1|+|\eta_2|-|\kappa|}  \nonumber \\
&=& [\beta_1 \beta_2]_{Q} \sum_{\kappa} \frac{\eta_2}{\kappa} (q^{\rho+\beta_1}) \frac{\eta_1}{\kappa}(q^{\rho+\beta_2}) Q^{|\eta_1|+|\eta_2|-|\kappa|} \,, \nonumber 
\ea
where in the last step, we have defined the pairing $[\cdot \cdot]_Q$.

\begin{figure}[h]
\begin{center}
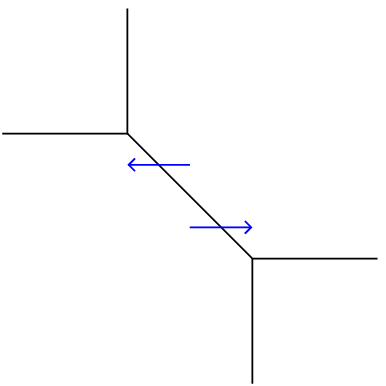
\end{center}
\caption{\small (-1,-1) curve. \label{m1m1curve}}
\end{figure}

\paragraph{Type $\{\beta_i \beta_j\}$: (-1,-1) curves:} The second type of pairing arises for (-1,-1) curves, as depicted in figure \ref{m1m1curve}.

The corresponding expression is
\ba \label{typeiisum}
C_{\at \ao \bo}  C_{\att \alpha_3 \beta_2} (-1)^{|\at|} Q^{|\at |} &=&   \bigg[ \bo\frac{\att}{\eta_1}(\beta_1) \frac{\ao}{\eta_1}(\bot)  \bigg] (-1)^{|\at|} Q^{|\at |} \bigg[\beta_2 \frac{\at}{\eta_2} (\beta_2) \frac{\alpha_3}{\eta_2} (\beta_2^t) \bigg]\,.
\ea
Note that here, the $\kappa(\alpha_2)$ dependence cancels between the two vertices, by $\kappa(\alpha^t)=-\kappa(\alpha)$. The framing factor $n$ vanishes. Again, this allows us to perform the sum over $\alpha_2$, utilizing (\ref{sumruletwo}). We obtain
\ba
\sum_\alpha \frac{\alpha^t}{\eta_1}(\beta_1) \frac{\alpha}{\eta_2}(\beta_2) (-Q)^{|\alpha|} &=& 
 \exp\left[-\sum_n \frac{Q^n}{n} \sf\left((q^{\rho+\beta_1})^n\right)\sf\left((q^{\rho+\beta_2})^n\right)\right] \label{defpair2} \\
 & & \sum_{\kappa} \frac{\eta_2^t}{\kappa} ((-Q)q^{\rho+\beta_1}) \frac{\eta_1}{\kappa}(q^{\rho+\beta_2}) (-Q)^{|\eta_1|} \nonumber \\
 &=& \{\beta_1 \beta_2 \}_{Q} \sum_{\kappa} \frac{\eta_2^t}{\kappa^t} (q^{\rho+\beta_1}) \frac{\eta_1^t}{\kappa}(q^{\rho+\beta_2}) (-Q)^{|\eta_1|+|\eta_2|-|\kappa|} \,. \nn
\ea
The last step defines the pairing $\{\cdot \cdot\}_Q= \frac{1}{[\cdot \cdot]_Q}$.

\subsection{Stringing the curves together}
\begin{figure}[htbp]
\begin{center}
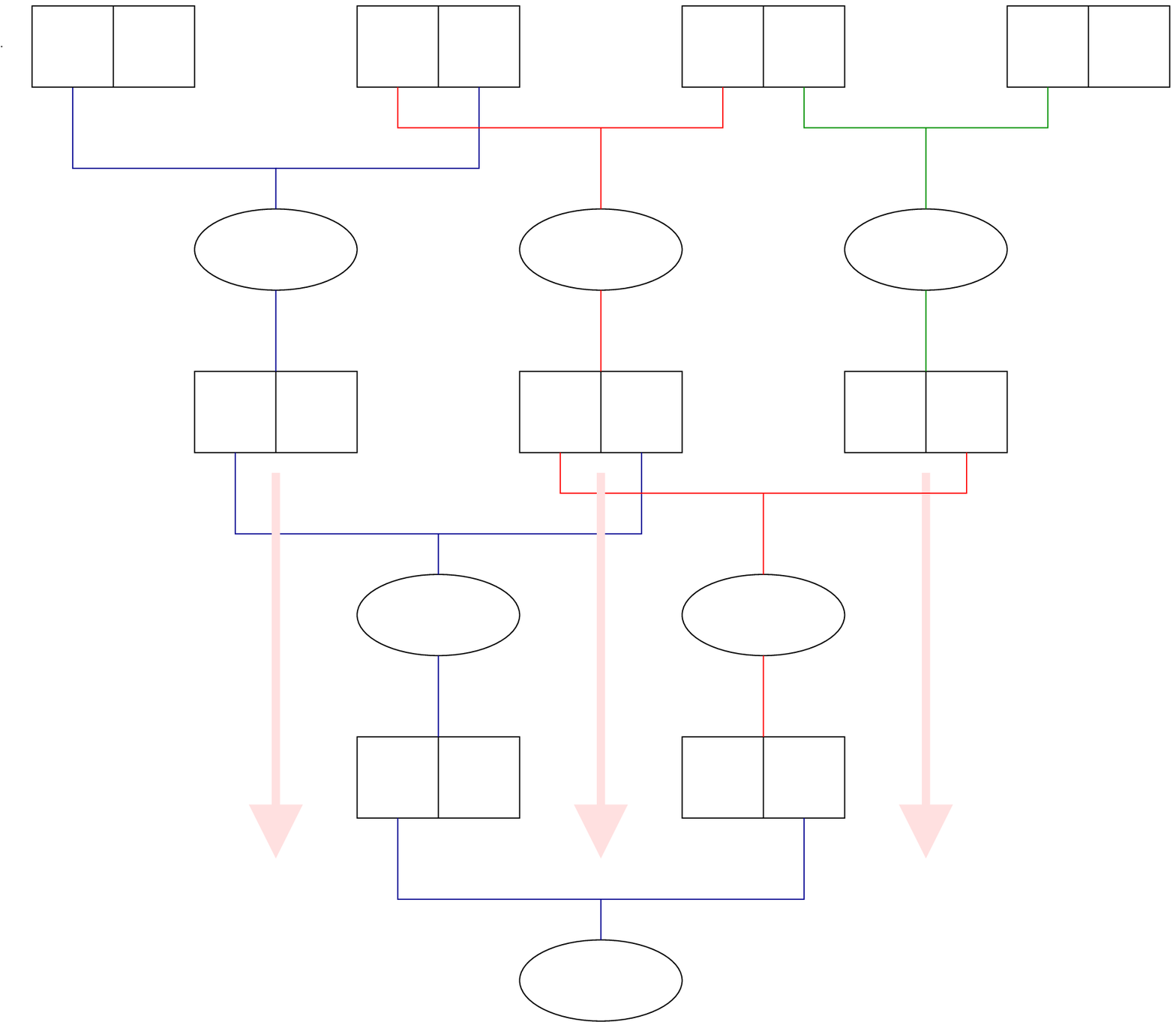
\end{center}
\caption{The gluing rules exemplified in a flow diagram.\label{calc}}
\end{figure}

The rules we have proposed in section 2 are easily proved by induction. Here, we wish to give some intuition as to how they arise. This is best done by considering an example. Let's therefore revisit figure \ref{example}.
We organize the calculation in a diagram, see figure \ref{calc}.
The following items should help explain and interpret the diagram.
\begin{itemize}
\item{The dominos in the first row correspond to the vertices of the web diagram describing the geometry. The connecting lines in between dominos indicate applications of the rules (\ref{sumruleone},\ref{sumruletwo}) for summing over skew Schur polynomials. The dominos in the second row arise after a depth one application of the summing rules, etc.}
\item{In each domino in the first row, either both of the top representations of the skew Schur functions (this terminology is to refer to the $\alpha$ in $\frac{\alpha}{\eta}$) are transposed or neither of them are. The former are vertices of type $C_{\alpha_{i+1} \alpha_i^t \beta_i}$ (type $A$), the latter of type $C_{\alpha_i^t \alpha_{i+1} \beta_i}$ (type $B$). In all the following rows, whether a skew Schur function carrying the argument $\beta_i^\cdot$  has transposed top representation or not depends on what is the case for the Schur function in the first row with argument $\beta_i^\cdot$. Hence, whether the pairing of $\beta_i^\cdot$ with $\beta_j^\cdot$ is of type $\{ \cdot \cdot \}$ or $[\cdot \cdot]$ is determined by the relative type of vertex the $\beta$'s descended from in the first row.}
\item{All  $\[\cdot \beta_i^\cdot\]$ pairings descend from the Schur function $\frac{\alpha_i^\cdot}{\eta_i}(\beta_i^\cdot)$ in the $i$-th vertex in the first row of the diagram, all $\[\beta_i^\cdot \cdot\]$ pairings from the Schur function $\frac{\alpha_{i+1}^\cdot}{\eta_i}(\beta_i^\cdot)$ in the same vertex. Hence, whether the correct entry is $\beta_i$ or $\beta_i^t$ again depends on whether the $i$-th vertex is of type $C_{\alpha_{i+1} \alpha_i^t \beta_i}$ (type $A$) or $C_{\alpha_i^t \alpha_{i+1} \beta_i}$ (type $B$).}
\item{The calculation terminates, because the first and last domino in each row contain a trivial skew Schur function $\frac{\bullet}{\eta}$, s.t. the sum over $\eta$ collapses.}
\item{The factors of $Q$ can be thought of as flowing along the connecting lines. Consider the second domino in the second line of the diagram. After applying (\ref{sumruletwo}), we obtain the $Q$ factor $Q_2^{|\eta_2| + |\eta_3| - |\kappa_2|}$. The diagram shows into which pairing these factors are incorporated. Note that the factors $Q_2^{|\eta_2|}$ and $Q_2^{|\eta_3|}$ enter into the next level of evaluation (sums over $\eta_2$ and $\eta_3$), whereas $Q_2^{-|\kappa_2|}$, as indicated by the arrows, enters in the level after next (the sum over $\kappa_2$).}
\item{Finally, a word on the factors $(-1)^{|\alpha|}$ that appear in sums that lead to the pairing $\{\cdot \cdot\}$, see (\ref{typeiisum}). One could combine these with their kin factors $Q^{|\alpha|}$, such that all $Q$'s associated to $(-1,-1)$ curves would come with a minus sign. The sign of the product of all $Q$ factors contributing to a pairing would then depend on whether an odd or even number of $(-1,-1)$ curves lie between the two vertices being paired. This of course is the criterion that distinguishes between the two pairings $\{\cdot \cdot\}$ and $[\cdot \cdot]$. Hence, this sign is taken into account correctly by incorporating it into the definition of $\{\cdot \cdot\}$ in (\ref{defpair2}).}
\end{itemize}

\subsection{Simplifying the two pairings}
The two pairings $\{\alpha \beta\}$ and $[\alpha \beta]$ are exponentials of the argument 
\ba \label{infsum}
\pm \left[ \sum_{n=1}^{\infty} \frac{Q^{n}}{n} \sf(q^{n(\rho + \alpha)}) \sf(q^{n(\rho + \beta)}) \right]  \,.
\ea
In this section, we perform the representation dependent part of this infinite sum. The calculation already appeared in \cite{Iqbal:2003ix,Iqbal:2003zz}. Our goal will be to write the product of Schur functions (up to a correction term) as a sum $\sum_{\rm finite} C_k q^{kn}$, which will allow us to subsume the infinite sum over $n$ in a logarithm.
First, let us take a closer look at the Schur polynomials. $\sf(x)=\sum_i x_i$, and hence,
\ba
\sf(q^{\rho + \alpha}) = \sum_{i=1}^{\infty} q^{\alpha_i - i + \frac{1}{2}} \,.
\ea
Apart from the finite number of terms involving the Young tableau $\alpha$, this is a geometric series,
\ba
\sf(q^{\rho + \alpha}) &=& \sqrt{q} \left( \sum_{i=1}^{\infty} q^{-i} + \sum_{i=1}^{d_\alpha} (q^{\alpha_i-i} - q^{-i}) \right)\\
&=&  \sqrt{q} \left( \frac{1}{q-1} + (q-1) \sum_{i=1}^{d_\alpha} q^{-i} \frac{q^{\alpha_i} - 1}{q-1} \right) \nn\\
&=&\frac{\sqrt{q}}{q-1} \left( 1 + (q-1)^2 \sum_{i=1}^{d_\alpha} q^{-i} \sum_{j=0}^{\alpha_i-1} q^j \right) \,.\nn
\ea
Here, $d_\alpha$ denotes the number of rows of $\alpha$. We now see the desired structure in the product $\sf(q^{n(\rho + \alpha)}) \sf(q^{n(\rho + \beta)})$ emerging, 
\ba
\sf(q^{\rho + \alpha}) \sf(q^{\rho + \beta})&=& \sum_k C_k(\alpha, \beta) q^k + \frac{q}{(1-q)^2} \,,
\ea
where
\ba \label{defck}
\sum_k C_k(\alpha, \beta)q^k &=& \frac{q}{(q-1)^2} \left( 1 + (q-1)^2 \sum_{i=1}^{d_\alpha} q^{-i} \sum_{j=0}^{\alpha_i-1} q^j \right) \left( 1 + (q-1)^2 \sum_{i=1}^{d_\beta} q^{-i} \sum_{j=0}^{\beta_i-1} q^j \right) \nonumber \\
& & - \frac{q}{(1-q)^2}  \,.
\ea
The infinite sum in (\ref{infsum}) can now be expressed in a more compelling form. For the minus sign which corresponds to the pairing $\{\cdot \cdot\}$, we obtain
\ba
-\sum_{n=1}^{\infty} \frac{Q^{n}}{n} \sf(q^{n(\rho + \alpha)}) \sf(q^{n(\rho + \beta)})  &=& - \sum_{n=1}^{\infty} \frac{Q^{n}}{n} \left[ \sum_k C_k(\alpha, \beta) q^{n k} + \frac{q^n}{(1-q^n)^2}  ) \right]  \\
&=&  \sum_k C_k(\alpha, \beta) \log (1- Q q^k) -  \sum_{n=1}^\infty \frac{(Qq)^n}{n(1-q^n)^2} \,. \nn
\ea
Recalling that $q=e^{i g_s}$, the remaining infinite sum takes a familiar form,
\ba
\sum_{n=1}^\infty \frac{(Qq)^n}{n(1-q^n)^2} &=& -\sum_{n=1}^{\infty} \frac{Q^n}{n(2 \sin( \frac{n g_s}{2}))^2} \,.
\ea
We hence obtain the tidy expression,
\ba
\{\alpha \beta\}_Q &=&  \prod_k  (1- Q q^k)^{ C_k(\alpha, \beta)} \exp \left[ \sum_{n=1}^\infty \frac{Q^n}{n(2 \sin( \frac{n g_s}{2}))^2} \right] \,.
\ea
When considering flops further below, we will need the relation between $\{\alpha \beta \}$ and $\{\alpha^t \beta^t\}$. The sum (\ref{defck}) satisfies the property
\ba
\sum_k C_k(\alpha, \beta)q^k &=&  \sum_k C_k(\alpha^t, \beta^t)q^{-k} \,.
\ea 
By the symmetry of the correction term $\frac{q}{(1-q)^2}$ under $q \rightarrow \frac{1}{q}$, it follows that
\ba
\{ \alpha^t \beta^t \}(q) &=& \{ \alpha \beta \}(\frac{1}{q}) \label{transposed}\,.
\ea 

\subsection{CS calculations in the light of the vertex} \label{CS}
\begin{figure}[htbp]
\begin{center}
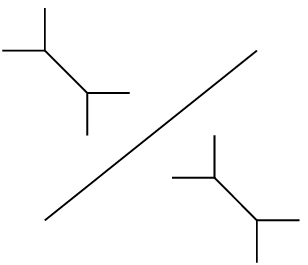
\end{center}
\caption{\small The relation between the CS 4-vertex and the topological vertex. \label{dcstovertex}}
\end{figure}

All genus results for the topological string on non-compact Calabi-Yau were originally obtained using Chern-Simons theory as the target space description of the open topological string, combined with open/closed duality via geometric transitions \cite{Diaconescu:2002sf,Diaconescu:2002qf,Aganagic:2002qg}. In the context of Chern-Simons theory, the natural object is a 4-vertex, the normalized expectation value of a Hopf link with gauge fields in representations $\alpha$ and $\beta$ on the two unknots. It is given by 
\ba \label{defW}
W_{\alpha \beta}(q,\lambda)=W_{\alpha}(q,\lambda)\,(q\lambda)^{|\beta|/2}\,S_{\beta}(E_{\alpha}(t))\,,
\ea
where
\ba \label{defE}
E_{\alpha}(t)=(1+\sum_{n=1}^{\infty}(\prod_{i=1}^{n}\frac{1-\lambda^{-1}q^{i-1}}{q^{i}-1})t^{n})\,(\prod_{j=1}^{d}\frac{1+q^{\alpha_{j}-j}t}{1+q^{-j}t})\,,
\ea
and 
\ba
W_{\alpha} = (q\lambda)^{|\alpha|/2} S_\alpha(E_\bullet(t))  \,.
\ea
We explain the relation of $S_\alpha(E(t))$ to the Schur functions $s_\alpha(x)$ of the previous sections in the appendix. The relevant fact for our purposes is that for $E(t) = \prod(1+x_i t)$, $S_\alpha(E(t)) = s_\alpha(x)$. To bring (\ref{defE}) into this form, we note that the first factor in that expression can be expressed as (see page 27, example 5 in \cite{Macdonald}),
\ba
1+\sum_{n=1}^{\infty}(\prod_{i=1}^{n}\frac{1-\lambda^{-1}q^{i-1}}{q^{i}-1})t^{n}&=& \prod_{i=0}^{\infty} \frac{1+\lambda^{-1}q^{i}t}{1+q^{i}t} \\
&=& \prod_{i=0}^{\infty} (1+\lambda^{-1}q^{i}t) \prod_{i=1}^{\infty} (1 + q^{-i}t) \,. \nn
\ea
Hence,
\ba
E_{\alpha}(t)= \prod_{i=0}^{\infty} (1+\lambda^{-1}q^{i}t) \prod_{j=1}^{\infty}(1+q^{\alpha_{j}-j}t)  \,.
\ea
In this equation, we have set $\alpha_j=0$ for $j>d$.
By absorbing the factor of $q^{|\alpha|/2}$ in (\ref{defW}) into the definition of $E$,
\ba
\tilde{E}_{\alpha}(t)=\prod_{i=1}^{\infty} (1+\lambda^{-1}q^{i-\frac{1}{2}}t) \prod_{j=1}^{\infty}(1+q^{\alpha_{j}-j+\frac{1}{2}}t)  \,,
\ea
we can now express $W_{\alpha \beta}$ in terms of ordinary Schur functions,
\ba
W_{\alpha \beta}&=& \lambda^{\frac{|\alpha|+|\beta|}{2}} S_{\alpha}(\tilde{E}_{\bullet}(t)) S_{\beta}(\tilde{E}_{\alpha}(t)) \\
&=& \lambda^{\frac{|\alpha|+|\beta|}{2}} s_{\alpha}(\lambda^{-1}q^{-\rho},q^\rho) s_\beta(\lambda^{-1}q^{-\rho}, q^{\rho+\alpha}) \,. \nn
\ea
Through a series of transformations, we can bring this expression into a form in which it can readily be related to the topological vertex,
\ba
W_{\alpha \beta}&=& \lambda^{\frac{|\alpha|+|\beta|}{2}}  \frac{\alpha}{\eta}(q^\rho) \, \eta^t \! \left((-\lambda)^{-1}q^{\rho}\right)  \frac{\beta}{\kappa}(q^{\rho+\alpha}) \,\kappa^t \! \left((-\lambda)^{-1}q^\rho\right) \label{cstovertex} \\
&=&  \lambda^{\frac{|\alpha|+|\beta|}{2}} \frac{\alpha}{\eta}(q^\rho) \, \frac{\frac{\delta}{\eta}(q^\rho) \, \delta^t(-Q q^\rho) }{ \{ \bullet \bullet \}_{Q}}  \frac{\beta}{\kappa}(q^{\rho+\alpha}) \frac{\frac{\gamma}{\kappa}(q^{\rho + \alpha^t}) \gamma^t(q^\rho) (-1)^{|\gamma|}Q^{|\gamma|}}{ \{\alpha^t \bullet\}_{Q}} \nonumber \\
&=&  \lambda^{\frac{|\alpha|+|\beta|}{2}} \alpha(q^\rho)\, \delta^t(q^{\rho + \alpha^t}) \frac{\delta(-Q q^\rho)}{ \{ \bullet \bullet\}_{Q} \{\alpha^t \bullet\}_{Q}} \frac{\beta}{\kappa}(q^{\rho+\alpha}) \frac{\gamma}{\kappa}(q^{\rho + \alpha^t}) \gamma^t(q^\rho) (-1)^{|\gamma|}Q^{|\gamma|} \nonumber \\
&=& \lambda^{\frac{|\alpha|+|\beta|}{2}} \frac{ C_{\beta^t \gamma \alpha} C_{\bullet \bullet \gamma^t} (-1)^{|\gamma|} Q^{|\gamma|}}{\{\bullet \bullet\}_{Q}}   \,, \nn
\ea
where $Q= \lambda^{-1}$.
In the course of these manipulations, we have used virtually all of the identities listed in the appendix. The relation (\ref{cstovertex}) between the CS 4-vertex and the topological vertex is depicted in figure \ref{dcstovertex}.

The mismatch of the factor $\lambda^{\frac{|\alpha|+|\beta|}{2}}$ was first noticed in \cite{Diaconescu:2002sf}.

\section{Applications}
\subsection{Flops}
A natural question to study is the behavior of the string partition function under flops of the target geometry. We can analyze this question for all geometries whose toric diagrams decompose into strips.

First, let us recall the behavior of the conifold under a flop. The instanton piece of the partition function is given by
\ba
Z_{conifold}^{inst}&=& \exp \left[ \sum_{n=1}^{\infty} \frac{Q^n}{n} \frac{1}{( 2 \sin \frac{n g_s}{2})^2} \right] \\
&=& \prod_{k=1}^{\infty} (1- Q q^{-k})^{k} \nonumber  \,.
\ea
In addition, the genus 0 and 1 free energy contain further polynomial dependence on $t$. The behavior of the polynomial contributions to $F_0$ under flops can be considered separately from that of the instanton contributions \cite{Witten:1993yc,Gopakumar:1998ki}. It turns out that the polynomial contribution to $F_1$, $\frac{1}{24}t$, is naturally considered together with the instanton contribution to the partition function. By analytic continuation, we obtain the partition function of the topological string on the flopped geometry from the partition function on the pre-flop geometry,
\ba
Q^{\frac{1}{24}} Z_{conifold}^{inst}(Q)  \rightarrow  Q^{-\frac{1}{24}} Z_{conifold}^{inst}(\frac{1}{Q})  \nonumber  \,.
\ea
To obtain the Gopakumar-Vafa invariants of the flopped geometry, we must now expand the RHS in the correct variable, $Q$,
\ba
Q^{-\frac{1}{24}} Z_{conifold}^{inst}(Q^{-1}) &=& Q^{-\frac{1}{24}} \prod_{k=1}^{\infty} (1-  \frac{q^{-k}}{Q})^{k}   \\
&=&Q^{-\frac{1}{24}} Q^{-\zeta(-1)} q^{-\zeta(-2)}\prod_{k=1}^{\infty} (Q q^{k} -1)^{k} \nn \\
&=&(-1)^{\zeta(-1)} Q^{\frac{1}{24}} \prod_{k=1}^{\infty} (1-Q q^{-k})^{k} \,. \nn
\ea
In the last line, we have used that $Z_{conifold}^{instanton}$ is invariant under $q \rightarrow \frac{1}{q}$, and $\zeta(-1)=-\frac{1}{12}$. We see that up to a phase, the partition function is invariant when analytically continued from $Q$ to $Q^{-1}$, and then re-expanded in powers of $Q$. It follows that the Gopakumar-Vafa invariants are in fact invariant under this flop.

\begin{figure}[h]
\begin{center}
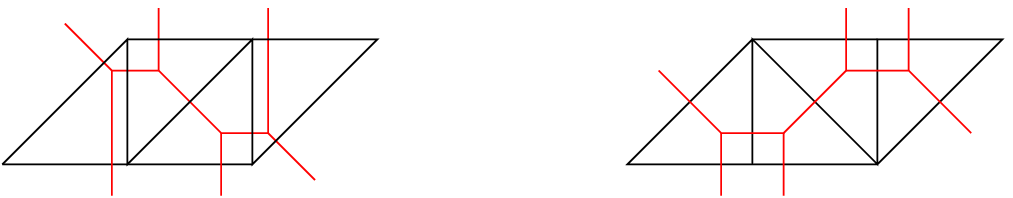
\vspace{0.3cm}
\caption{$(-1,-1)(-1,-1)_Q(-1,-1)$ flopped to $(-2,0)(-1,-1)_{Q^{-1}}(-2,0)$. \label{flopcase1}}
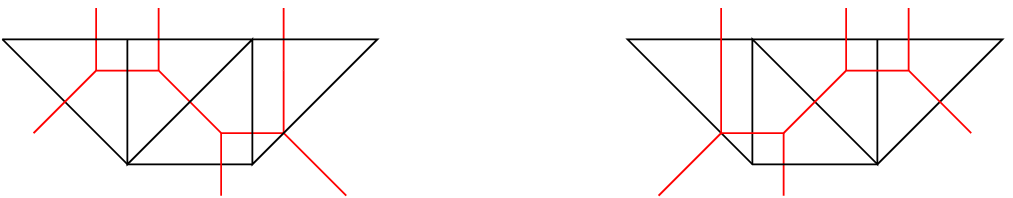
\vspace{0.3cm}
\caption{$(-2,0)(-1,-1)_Q(-1,-1)$ flopped to $(-1,-1)(-1,-1)_{Q^{-1}}(-2,0)$.\label{flopcase2}}
\end{center}
\end{figure}

\begin{figure}
\begin{center}
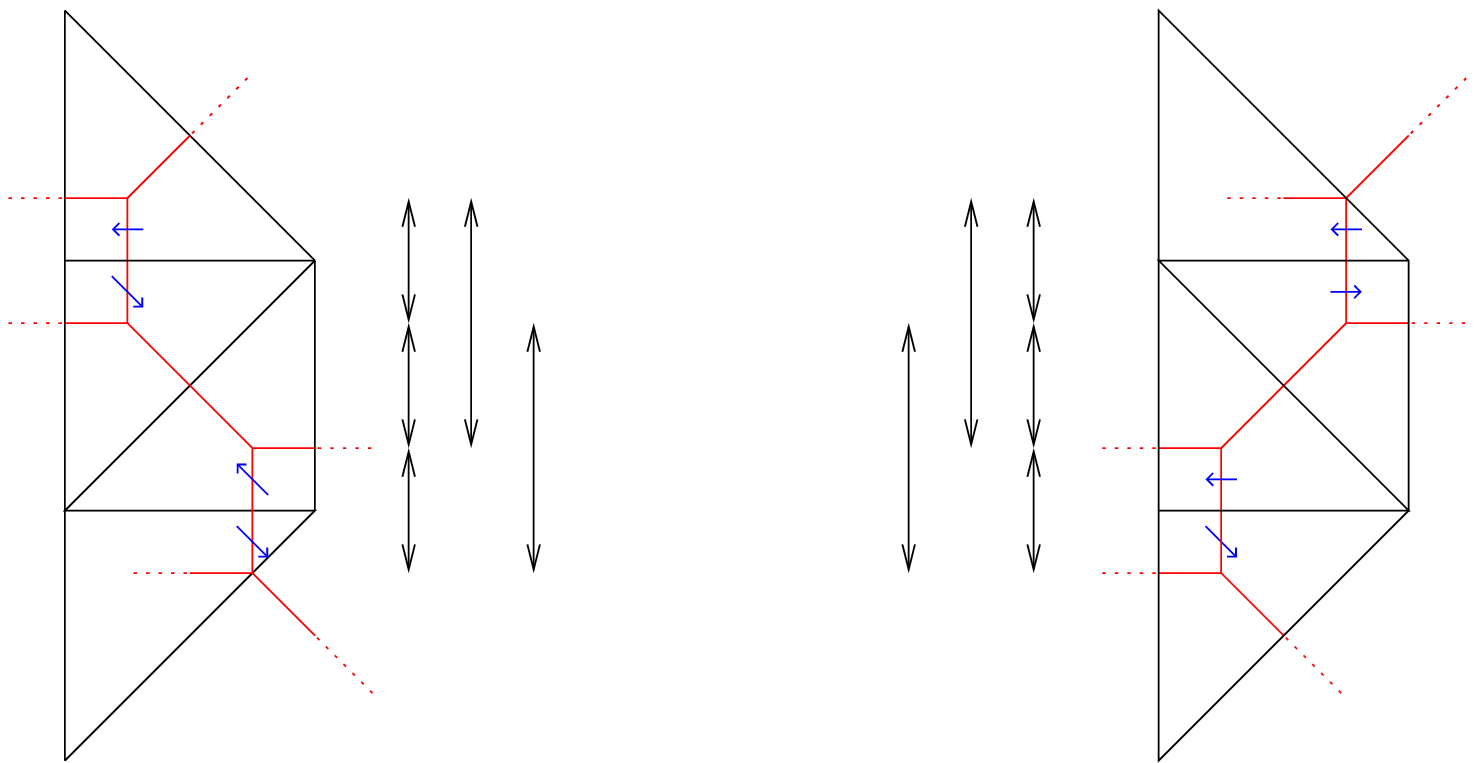
\end{center}
\caption{Embedding the previous figure in a larger geometry, with relevant framings indicated. \label{floptopbottom}}
\end{figure}

Now let's turn to flops on the strip. The normal bundles of the curves neighboring the flopped curve are affected by the flop. On the strip, two geometries are to be distinguished: the $(-1,-1)$ curve to be flopped is connected along the strip to a $(-1,-1)$ curve on both sides (figure \ref{flopcase1}), or to a $(-1,-1)$ curve on one side, and a $(-2,0)$ curve on the other (figure \ref{flopcase2}). After the flop, the $(-1,-1)$ curves become $(-2,0)$ curves in the former case, and the $(-1,-1)$ and $(-2,0)$ curves are swapped in the latter.
We will consider the first case in detail. The second works out in exact analogy. Before the flop, we have three $(-1,-1)$ curves. The first vertex is $C_{\bullet \alpha_1 \beta_1}$, hence of type $B$, from which we can determine the type of all vertices to be $BABA$, yielding
\ba
\frac{ \{\beta_1^t \beta_2^t\}_{Q_1} \{\beta_1^t \beta_4^t\}_{Q_1 Q_2 Q_3} \{\beta_2 \beta_3\}_{Q_2} \{\beta_3^t \beta_4^t\}_{Q_3}}{ \{\beta_1^t \beta_3\}_{Q_1 Q_2} \{\beta_2 \beta_4^t\}_{Q_2 Q_3}} \,. \label{preflop}
\ea
After the flop, note the important fact that the ordering of the vertices does not coincide with the ordering of the indices of the external Young tableaux the vertices carry. Assembling the data required to apply our rules in short hand: \{$(-2,0)$, $(-1,-1)$, $(-2,0)$ curves, first vertex of type $B$\} $\rightarrow BBAA$, we obtain
\ba
\frac{ \{\beta_1^t \beta_2^t\}_{\Qn_1 \Qn_2} \{\beta_1^t \beta_4^t\}_{\Qn_1 \Qn_2 \Qn_3} \{\beta_3^t \beta_2^t\}_{\Qn_2} \{\beta_3^t \beta_4^t\}_{\Qn_2 \Qn_3}}{ \{\beta_1^t \beta_3\}_{\Qn_1} \{\beta_2 \beta_4^t\}_{\Qn_3}} \,. \label{postflop}
\ea
To compare these two expressions, we can express the post-flop expression in terms of the pre-flop K\"ahler parameters, and then reexpand as in the case of the conifold considered above. The identification of the K\"ahler parameters that we propose is obtained by matching corresponding curves before and after the flop. Considering e.g. the $(-1,-1)$ curves before and after the flop, we obtain the relations
\ba
Q_1 &=& \Qn_1 \Qn_2 \,, \nonumber \\
Q_2 &=& \frac{1}{\Qn_2} \,, \nonumber \\
Q_3 &=& \Qn_2 \Qn_3 \,, \nonumber
\ea
which are consistent with the identification we obtain by considering $(-2,0)$ curves,
\ba
Q_1 Q_2 &=& \Qn_1 \,, \nn \\
Q_2 Q_3 &=& \Qn_3 \,. \nn
\ea
This identification of K\"ahler parameters is to be contrasted to the naive substitution $Q_2 \mapsto Q_2^{-1}$ for each curve whose K\"ahler parameter has $Q_2$ dependence.
Upon making these substitutions, the only factor in (\ref{postflop}) which must be reexpanded is $\{\beta_2^t \beta_3^t\}_{Q_2^{-1}}$. Using relation (\ref{transposed}) and the definition of the pairing, we obtain (dropping indices and renaming tableaux for ease of notation),
\ba
\{\alpha^t \beta^t\}_{Q^{-1}}(q) &=& \{\alpha \beta\}_{Q^{-1}}(\frac{1}{q})  \label{floppin} \\
&=&  \prod_k  (1- Q^{-1} q^{-k})^{ C_k(\alpha, \beta)} \exp \left[ \sum_{n=1}^\infty \frac{Q^{-n}}{n(2 \sin( \frac{n g_s}{2}))^2} \right] \nonumber \\
&=&  \prod_k (Q^{-1}q^{-k})^{C_k(\alpha ,\beta)} (Q q^{k}-1)^{C_k(\alpha, \beta)} (-Q)^{\frac{1}{12}}\exp \left[ \sum_{n=1}^\infty \frac{Q^{n}}{n(2 \sin( \frac{n g_s}{2}))^2} \right] \nonumber \\
&=&  Q^{-|\alpha|-|\beta|}q^{-\frac{\kappa(\alpha)+\kappa(\beta)}{2}}\prod_k (Q q^{k}-1)^{C_k(\alpha, \beta)} (-Q)^{\frac{1}{12}}\exp \left[ \sum_{n=1}^\infty \frac{Q^{n}}{n(2 \sin( \frac{n g_s}{2}))^2} \right] \nonumber \\
&=& (-Q)^{-|\alpha|-|\beta|}q^{-\frac{\kappa(\alpha)+\kappa(\beta)}{2}} (-Q)^{\frac{1}{12}} \{\alpha \beta\}_Q (q)\,, \nn
\ea
where we have used \cite{Iqbal:2003ix}
\ba 
\sum_k C_k(\alpha,\beta) &=& |\alpha|+|\beta| \,, \\
\sum_k k\, C_k(\alpha,\beta) &=& \frac{\kappa(\alpha)+\kappa(\beta)}{2}\,.
\ea
This almost coincides with the partition function (\ref{preflop}) for the pre-flop geometry. To interpret the coefficients, let's include the two curves that are connected to the flopped curve via a sum over the representations $\beta_2$ and $\beta_3$ into our considerations. Figure \ref{floptopbottom} shows an example.
The $\beta_2$ and $\beta_3$ dependent factors from the upper and lower part of the diagram before the flop are
\ba
Q_t^{|\beta_2|} (-1)^{n_t |\beta_2|} q^{-\frac{n_t \kappa(\beta_2)}{2}}
Q_b^{|\beta_3|} (-1)^{n_b |\beta_3|} q^{-\frac{n_b \kappa(\beta_3)}{2}} = Q_t^{|\beta_2|} (-1)^{|\beta_2|} q^{-\frac{\kappa(\beta_2)}{2}}
Q_b^{|\beta_3|}  \label{precoef} \,,
\ea
where $n_t = (-1, 0) \wedge (1, -1) = 1$ and $n_b = (1, -1) \wedge (-1, 1) = 0$. After the flop, we have
\ba
\Qn_t^{|\beta_2|} (-1)^{\tilde{n}_t |\beta_2|} q^{-\frac{\tilde{n}_t \kappa(\beta_2)}{2}}
\Qn_b^{|\beta_3|} (-1)^{\tilde{n}_b |\beta_3|} q^{-\frac{\tilde{n}_b \kappa(\beta_3)}{2}}=\Qn_t^{|\beta_2|} \Qn_b^{|\beta_3|} (-1)^{- |\beta_3|} q^{\frac{ \kappa(\beta_3)}{2}} \label{postcoef} \,,
\ea
with $\tilde{n}_t = (-1 ,0) \wedge (1, 0) = 0$ and $\tilde{n}_b = (1, -1) \wedge (-1, 0) = -1$. Combining (\ref{postcoef}) with the coefficients from (\ref{floppin}), we obtain (\ref{precoef}),
\ba
 (-Q_2)^{-|\beta_2|-|\beta_3|}q^{-\frac{\kappa(\beta_2)+\kappa(\beta_3)}{2}} \times \Qn_t^{|\beta_2|} \Qn_b^{|\beta_3|} (-1)^{- |\beta_3|} q^{\frac{ \kappa(\beta_3)}{2}} &=&  Q_t^{|\beta_2|} Q_b^{|\beta_3|} (-1)^{ |\beta_2|} q^{-\frac{ \kappa(\beta_2)}{2}} \,.
\ea
We see that the coefficients in (\ref{floppin}) are exactly those needed to maintain invariance of the Gopakumar-Vafa invariants under flops.

This result continues to hold for the situation depicted in figure \ref{flopcase2}, as well as the other possible completions of the lines carrying the Young tableaux $\beta_2$ and $\beta_3$ by curves of type $(-2,0)$ or $(-1,-1)$.

Can we conclude that Gopakumar-Vafa invariants for toric Calabi-Yau are invariant under flops in general? The above arguments were valid for geometries which contained only $(-1,-1)$ and $(-2,0)$ curves. For the general case, the topological string partition function
\ba
Z=\exp \left[ \sum_{\vec{n}}\sum_k \frac{n^g_{\vec{n}}\; Q^{k\vec{n}} }{k(2 \sin(\frac{k g_s}{2}))^{2-2g}} \right]
\ea
can be put in the product form \cite{Hollowood:2003cv,Katz:2004js}
\ba
Z=\prod_{\vec{n}} \left( \prod_{k=1}^{\infty} (1- q^k Q_{\vec{n}})^{k n^0_{\vec{n}}}  \prod_{g=1}^{\infty} \prod_{k=0}^{2g-2} (1 - q^{g-1-k}Q^{\vec{n}} )^{(-1)^{k+g} n^{g}_{\vec{n}} {2g-2 \choose k}}\right)    \,,
\ea
where $\vec{n}$ encodes the classes of the various holomorphic curves relative to the basis specified by the $Q_i$. By the same argument as above, the only factor which needs to be re-expanded after expressing the post-flop partition function in pre-flop variables is the one counting contributions from just the flopped curve. As above, this factor is invariant, up to a coefficient, under this operation. However, it remains to argue that for the curves we identify before and after the flop, say $\vec{n}$ and $\vec{\tilde{n}}$, the relation $n^g_{\vec{n}} = n^g_{\vec{\tilde{n}}}$ holds. For the case of geometries with only $(-1,-1)$ and $(-2,0)$ curves, this equality followed from the comparison of (\ref{preflop}) and (\ref{postflop}), and the interpetation of the coefficients in (\ref{floppin}).

Note that after an appropriate number of blowups and flops, any toric CY can be decomposed into the strips we have been considering in this paper. Hence, if Gopakumar-Vafa invariants indeed prove to be invariant under flops in general, the vertex calculations on a strip performed in this paper become relevant for any toric CY.

\subsection{Geometric Engineering}
\begin{figure}
\begin{center}
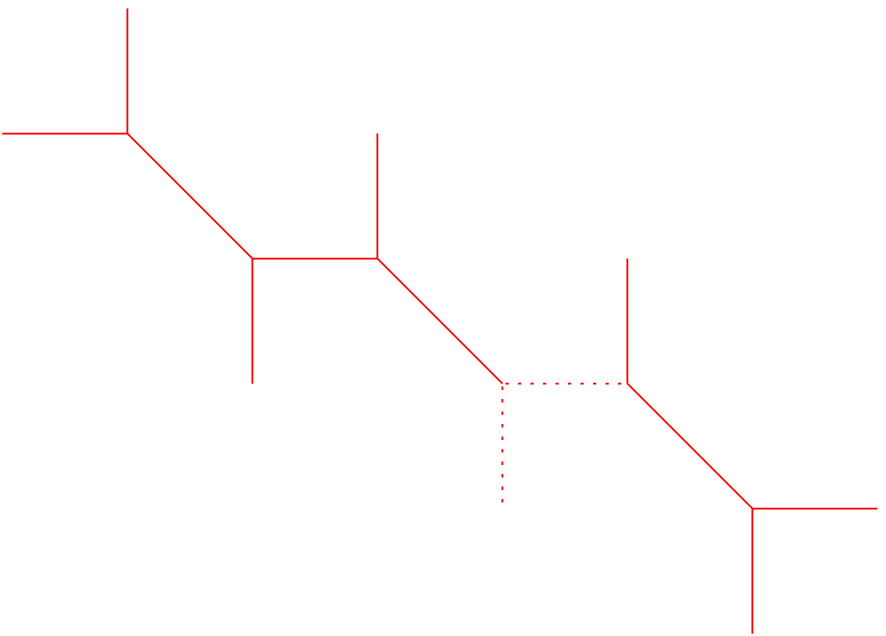
\end{center}
\caption{The building block for gauge theories with matter.\label{bblock}}
\end{figure}

A natural physical playground for the formalism developed in this paper is in the context of geometric engineering of $\N=2$ gauge theories by compactification on local CY. The toric geometries that give rise to linear chains of $U(N)$ gauge groups (i.e. theories with product gauge group with $U(N_i)$ factors and bifundamental matter between adjacent gauge groups) can be decomposed into the strips we consider here. Thanks to Nekrasov's construction, the full string partition function (vs. only its field theory limit) can be extracted from such gauge theories. The basic building block of the geometry which engineers such gauge theories is the triangulation of the strip given by \includegraphics{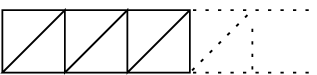}. The corresponding web
diagram is shown in figure \ref{bblock}. 
All of the curves in figure (\ref{bblock}) are (-1,-1) curves, and the first vertex is of type $A$, hence we have an alternating succession of vertices $ABABAB\ldots$. By the rules derived above, we obtain
\ba 
K^{\alpha_{1}\cdots \alpha_{N}}_{\beta_{1}\cdots\beta_{N}} &=&\frac{{\cal
K}^{\alpha_{1}\cdots \alpha_{N}}_{\beta_{1}\cdots\beta_{N}}}{
{\cal
K}^{\bullet \cdots \bullet}_{\bullet \cdots \bullet}} \label{basic} \\
&=&{\cal W}_{\alpha_{1}}{\cal W}_{\beta_{1}}\cdots {\cal
W}_{\alpha_{N}}{\cal W}_{\beta_{N}}\,\times
\\\nonumber
&&\prod_{k}\frac{\prod_{i\leq
j}(1-q^{k}Q_{\alpha_{i}\beta_{j}})^{C_{k}(\alpha_{i},\beta_{j}^{t})}
\prod_{i<j}(1-q^{k}Q_{\beta_{i}\alpha_{j}})^{C_{k}(\beta_{i}^t,\alpha_{j}^{t})}}
{\prod_{i<j}(1-q^{k}Q_{\alpha_{i}\alpha_{j}})^{C_{k}(\alpha_{i},\alpha_{j}^{t})}
(1-q^{k}Q_{\beta_{i}\beta_{j}})^{C_{k}(\beta_{i}^t,\beta_{j})}} \nn 
\ea
The K\"ahler parameters $Q_{\alpha,\beta}$ are given by
\ba Q_{\alpha_{i}\alpha_{j}}&=&Q_{ij}\\\nonumber
Q_{\alpha_{i}\beta_{j}}&=&Q_{ij}Q_{m,j}\\\nonumber
Q_{\beta_{i}\alpha_{j}}&=&Q_{ij}Q_{m,i}^{-1}\,,\\\nonumber
Q_{\beta_{i}\beta_{j}}&=&Q_{ij}Q_{m,i}^{-1}Q_{m,j}\,, \ea where
$Q_{ij}=\prod_{k=i}^{j-1}Q_{m,k}Q_{f,k}$.

${\cal K}^{\alpha_{1} \cdots \alpha_{N}}_{\beta_{1}\cdots\beta_{N}}$ is the building block for the partition function $Z_{Nekrasov}$ of ${\cal N}=2$ gauge theories
with product gauge groups and bi-fundamental matter. The partition function is given by products of ${\cal
K}^{\alpha_{1},\cdots, \alpha_{N}}_{\beta_{1},\cdots,\beta_{N}}$,
in the field theory limit, summed over $\alpha_{i},\beta_{i}$, where the field theory limit of ${\cal K}^{\alpha_{1},\cdots,
\alpha_{N}}_{\beta_{1},\cdots,\beta_{N}}$ is given by
\ba\nonumber
K^{\alpha_{1},\cdots,\alpha_{N}}_{\beta_{1},\cdots,\beta_{N}} \rightarrow{\cal
L}_{\alpha_{1}}{\cal L}_{\beta_{1}}\cdots {\cal
L}_{\alpha_{N}}{\cal L}_{\beta_{N}}\prod_{k}\frac{\prod_{i\leq
j}(a_{i}+m_{j}+k\hbar)^{C_{k}(\alpha_{i},\beta^{t}_{j})}\prod_{i<j}(a_{j}-m_{i}+k\hbar)^{C_{k}(\beta_{i}^{t},\alpha_{j}^{t})}}
{\prod_{i<j}(a_{ij}+k\hbar)^{C(\alpha_{i},\alpha_{j}^{t})}(m_{j}-m_{i}+k\hbar)^{C_{k}(\beta_{i}^{t},\beta_{j})}}
\ea with \ba \nonumber {\cal L}_{\alpha}=\mbox{lim}_{q\rightarrow
1}(q-1)^{|\alpha|}{\cal W}_{\alpha}\,.\ea

\begin{figure}
\begin{center}
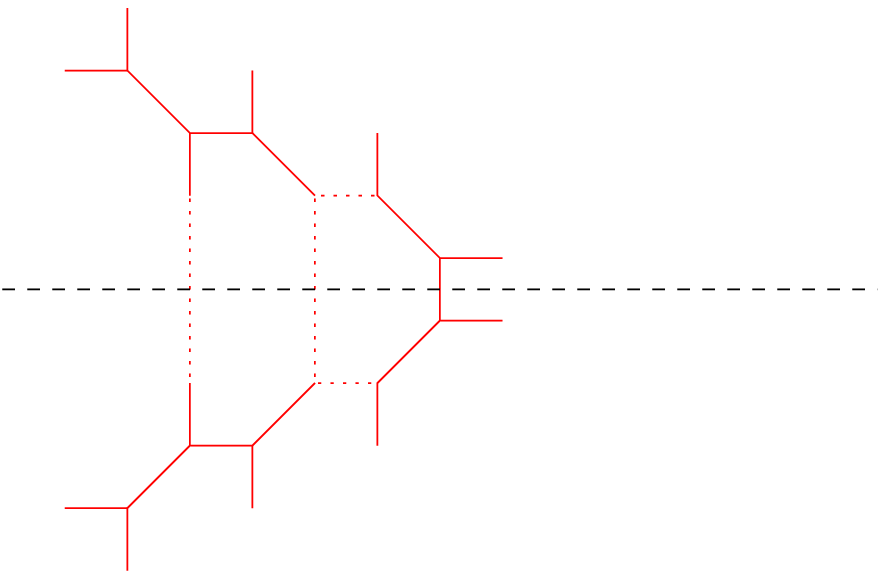
\end{center}
\caption{Two strips glued together to engineer $U(N)$ with matter.\label{sunwithmatter}}
\end{figure}

As an example, consider the CY in figure (\ref{sunwithmatter}) which
can be used to engineer $U(N)$ with $N_{f}=2N$.
The partition function is obtained by gluing two ${\cal K}$ type expressions together. Note however that in the upper strip, the order of vertices is $BABABA\ldots$. A moment's thought teaches us that $\K_{upper}(\alpha,\beta) = \K_{lower}(\alpha^t,\beta^t)$. Hence,
\ba
Z&=&\sum_{\alpha_{1},\cdots,\alpha_N} Q_{b}^{|\alpha_{1}|+\cdots+|\alpha_{N}|}{\cal
K}^{\alpha_{1} \cdots \alpha_{N}}_{\bullet \cdots \bullet}(Q){\cal
K}^{\bullet \cdots \bullet}_{\alpha_{1} \cdots \alpha_{N}}(\Qn)  \\
&=&\sum_{\alpha_{1},\cdots,\alpha_N}Q_{b}^{|\alpha_{1}|+\cdots+|\alpha_{N}|}{\cal
W}_{\alpha_{1}}^{2}\cdots {\cal W}_{\alpha_{N}}^{2}\times
\prod_{k}\prod_{i=1}^{N}(1-q^{k}Q_{m,i})^{C_{k}(\alpha_{i},\bullet)}
(1-q^{k}\tilde{Q}_{m,i})^{C_{k}(\bullet,\alpha_{i})}
\times \nn \\
&&\prod_{i<j}\frac{(1-q^{k}Q_{ij}Q_{m,j})^{C_{k}(\alpha_{i})}
(1-q^{k}Q_{ij}\tilde{Q}_{m,j}^{-1})^{C_{k}(\alpha^{t}_{i})}
(1-q^{k}Q_{ij}Q_{m,i}^{-1})^{C_{k}(\alpha_{j}^{t})}(1-q^{k}Q_{ij}\tilde{Q}_{m,i})^{C_{k}(\alpha_{j})}}
{(1-q^{k}Q_{ij})^{2C_{k}(\alpha_{i},\alpha_{j}^{t})}} \,. \nn
\label{PF}\ea
Defining
\ba
Q_{ij}&=&e^{-\beta(a_{i}-a_{j})}\,,\\\nonumber
Q_{m,i}&=&e^{-\beta(a_{i}+m_{i})}\,,\\\nonumber
\tilde{Q}_{m,i}&=&e^{-\beta(a_{i}+m_{i+N})}\,,\\\nonumber
q&=&e^{-\beta\hbar}\,, \ea
the field theory limit is given by $\beta \rightarrow 0$. In this limit, (\ref{PF}) yields \ba {\cal
Z}=\sum_{\alpha_{1,\cdots,N}}Q_{b}^{|\alpha_{1}|+\cdots+|\alpha_{N}|}
{\cal
Z}^{(0)}_{\alpha_{1}\cdots\alpha_{N}}\prod_{k}\prod_{i,j}(a_{i}+m_{j}+k\hbar)^{C_{k}(\alpha_{i})}(a_{i}+m_{j+N}+k\hbar)^{C_{k}(\alpha_{i})} \,,
\ea which is Nekrasov's partition function (equation (1.8) in
\cite{Nekrasov:2003af}) for $N_{f}=2N$ after using the identities
given in \cite{Iqbal:2003zz}.

\section{Conclusion}
How to move off the strip? We saw that an obstacle to taking a turn off the strip was performing the sums (\ref{sumruleone}) and (\ref{sumruletwo}) with factors of type $q^{\kappa(\alpha)/2}$ included in the sum over $\alpha$. This obstacle does not appear insurmountable, and efforts are underway to evaluate such sums. With them, all sums in the expression for the topological partition function of toric manifolds whose web diagram consists of a closed loop with external lines attached could be performed.
To go further, one would need to perform sums over the Young tableaux which are the arguments of the Schur functions which appear in the topological vertex. 

\section*{Acknowledgements}
We would like to thank Cumrun Vafa for many valuable discussions. AK would also like to thank Mina Aganagic, Paul Aspinwall, and Bogdan Florea for helpful conversations, as well as the Simons Workshop in Mathematics and Physics and the Harvard High Energy Physics Group for a very productive atmosphere in which part of this work was completed.

The work of AK was supported by the U.S. Department of Energy under contract number DE-AC02-76SF00515.

\appendix

\section{Getting to know Schur functions}
Since Schur functions feature prominently in this text, we wish to briefly present them in their natural habitat of symmetric functions in this appendix. Readers who wish to learn more are referred to, e.g., \cite{Macdonald, Fulton}.

Schur polynomials $s_\lambda(x_1,\ldots,x_k)$ present a basis of the symmetric polynomials in $k$ variables. They arise in representation theory as the characters of the Schur functor. Two perhaps more intuitive choices of basis for the symmetric polynomials are the following. The complete symmetric polynomials $h_r$ are defined as the sum of all monomials in $k$ variables of degree $r$, e.g. for $k=2$, $r=2$, $h_2=x_1^2+x_2^2+x_1 x_2$, and the elementary symmetric polynomials $e_r$ as the sum of all monomials of degree $r$ in distinct variables, e.g. $e_2=x_1 x_2$. To a Young tableaux $\lambda$ with at most $k$ rows, one now introduces the polynomial $h_\lambda = h_{\lambda_1} \cdots h_{\lambda_n}$, where $\lambda_i$ denotes the number of boxes in the $i$-th row of $\lambda$, and likewise, to a Young tableaux $\mu$ such that $\mu^t$ has at most $k$ rows, $e_\mu = e_{\mu_1} \cdots e_{\mu_n}$. Both sets $\{h_\lambda\}$, $\{e_\mu\}$ comprise a basis for symmetric polynomials. The Schur polynomials can be expressed in terms of these, using the so-called determinantal formulae,
\ba
s_{\lambda}&=& | h_{\lambda_i + j-i}| \\
&=& | e_{\lambda^t_i + j -i}|  \,. \label{JT}
\ea
The skew Schur polynomials, which we introduced in the text via their relation to the ordinary Schur polynomials, $s_{\lambda/\mu}(x) = \sum_\nu c^\lambda_{\mu \nu} s_\nu(x)$, also satisfy determinantal identities,\footnote{The difference of two Young tableaux (performed row-wise), as it appears in the determinantal formulae (\ref{detss1}) and (\ref{detss2}), is also called a skew Young tableau, hence the name {\it skew} Schur polynomials.}
\ba
s_{\lambda/\mu}&=& | h_{\lambda_i - \mu_j + j-i}| \label{detss1}\\
&=& | e_{\lambda^t_i -\mu^t_j + j -i}|  \,. \label{detss2}
\ea
The generating function for the elementary symmetric functions $e_i$ is $\prod (1+x_i t)$, i.e. the coefficient of $t^i$ in this power series is the $i$-th elementary symmetric function $e_i$. We now define the functions $S_\lambda(E(t))$, which we encountered in section (\ref{CS}), in accordance with the determinantal formula (\ref{JT}), where $e_i$ is replaced by the coefficient of $t^i$ in the power series $E(t)$. Clearly, for $E(t)= \prod (1+x_i t)$, $S_\lambda(E(t)) = s_\lambda(x)$.

Next, we collect the identities for the Schur functions we use in the text.
\ba
\sum_\alpha s_{\alpha/\eta_1}(x)s_{\alpha/\eta_2}(y)&=&  \prod_{i,j} (1-x_i y_j)^{-1} \sum_\kappa s_{\eta_2/\kappa}(x)s_{\eta_1/\kappa}(y) \,, \\
\sum_\alpha s_{\alpha^t/\eta_1}(x)s_{\alpha/\eta_2}(y) &=&  \prod_{i,j} (1+x_i y_j) \sum_\kappa s_{\eta_2^t/\kappa^t}(x)s_{\eta_1^t/\kappa}(y) \,,\\
\sum_\alpha s_{\alpha/\eta}(x) s_{\eta}(y) &=& s_\alpha(x,y) \,,\\
s_\alpha(q^{\rho + \beta}) &=& (-1)^{|\alpha|} s_{\alpha^t}(q^{-\rho-\beta^t}) \,,\\
s_\alpha(q^\rho) &=& q^{\frac{\kappa(\alpha)}{2}} s_{\alpha^t}(q^\rho) 
\,.
\ea 
By invoking the cyclicity of the vertex, we further obtain
\ba
s_{\alpha}(q^\rho) s_{\beta}(q^{\rho+\alpha}) &=& s_{\beta}(q^\rho) s_{\alpha}(q^{\rho+\beta})  \,, \\
q^{\frac{\kappa_\beta}{2}} s_{\alpha}(q^\rho) s_{\beta^t}(q^{\rho + \alpha^t}) &=& \sum_\eta s_{{\alpha}/{\eta}}(q^\rho) s_{{\beta}/{\eta}}(q^\rho) \,.
\ea

In our applications, we need to work within the ring of symmetric functions with countably many independent variables. This ring can be obtained from the rings of symmetric polynomials in finitely many variables via an inverse limit construction \cite{Macdonald}.

\end{document}